# Application of Superhalogens in the Design of Organic Superconductors

Ambrish K. Srivastava,*[a] Abhishek Kumar,[b] Sugriva N. Tiwari,[a] and Neeraj Misra[b]

Dedicated to Prof. K. Bechgaard, who passed away on March 7, 2017

**Abstract:** Bechgaard salts, $(TMTSF)_2X$ (TMTSF = tetramethyl tetraselenafulvalene and X = complex anion), form the most popular series of organic superconductors. In these salts, TMTSF molecules act as super-electron donor and X as acceptor. We computationally examine the electronic structure and properties of X in commonly used $(TMTSF)_2X$ (X = $NO_3$, $BF_4$, $ClO_4$, $PF_6$) superconductors and notice that they belong to the class of superhalogens due to their higher vertical detachment energy than halogen anions. This prompted us to choose other superhalogens such as X = $BO_2$, $BH_4$, $B_2F_7$, $AuF_6$ and study their $(TMTSF)_2X$ complexes. Our findings suggest that these complexes behave more or less similar to those of established $(TMTSF)_2X$ superconductors, particularly for X = $BO_2$ and $B_2F_7$. We, therefore, believe that the concept of superhalogen can be successfully applied in the design of novel organic superconductors.

Organic superconductors (OSCs) are of particular interest due to their high anisotropy and other intriguing properties. W. A. Little [1] proposed for the first time in 1964 the possibility of synthesis of organic polymers, which might exhibit superconductivity. K. Bechgaard [2] discovered the first organic superconductor, $(TMTSF)_2PF_6$ in 1980 (TMTSF = tetramethyl tetraselenafulvalene), which led to the synthesis of a series of related organic compounds, known as Bechgaard salts [3]. In this series, $(TMTSF)_2ClO_4$ is particularly interesting due to its superconducting property at ambient pressure [4]. These materials are considered quasi one-dimensional due to the fact that superconduction can only occur along a single axis. The Fabre salts form another series of compounds belonging to this class, which are composed of tetramethyl tetrathiafulvalene (TMTTF). The quasi two-dimensional materials such as bisethylenedithio tetrathiafulvalene (BEDT-TTF) series and three-dimensional alkali metal doped fullerenes based organic superconductors are also well studied [5, 6]. The alkali-doped fullerene $RbCs_2C_{60}$ is known to possess the highest critical temperature of 33 K at ambient pressure [7]. Despite these facts, Bechgaard salts have many other properties that make them particularly interesting. For instance, these salts can easily undergo any phase by varying both temperature and pressure.

In $(TMTSF)_2X$ (where $X^-$ = complex anion), TMTSF serves as electron donor and X behaves as electron acceptor, thus forming a charge transfer complex with metal like characteristics.

Although much attention has been paid to electron donors for OSCs [8], the studies on the role of X are relatively scarce. Here, we focus on the electronic structure and properties of complex anions ($X^-$) and notice that they all belong to a special class of species, known as superhalogen. Superhalogen, proposed by Gutsev and Boldyrev [9] in 1981 and experimentally confirmed by Wang et al. [10] in 1999, are the species with higher electron affinity (EA) or vertical detachment energy (VDE) than those of halogen. Superhalogens have been continuously studied [11] due to their possible applications in a variety of fields [12]. This provides us an opportunity to analyze other superhalogens for their possible role in the design of new materials for organic superconductors.

We first analyze the molecular structure and superhalogen nature of complex anions ($X^-$) commonly employed in $(TMTSF)_2X$ series of organic superconductors such as $NO_3^-$, $BF_4^-$, $ClO_4^-$, $PF_6^-$, etc. as displayed in Fig. 1. The structures of $NO_3^-$ and $PF_6^-$ are trigonal planar and octahedral with the bond lengths of 1.26 Å and 1.65Å, respectively, whereas both $BF_4^-$ and $ClO_4^-$ are tetrahedral with the bond lengths of 1.42 Å and 1.50 Å, respectively.

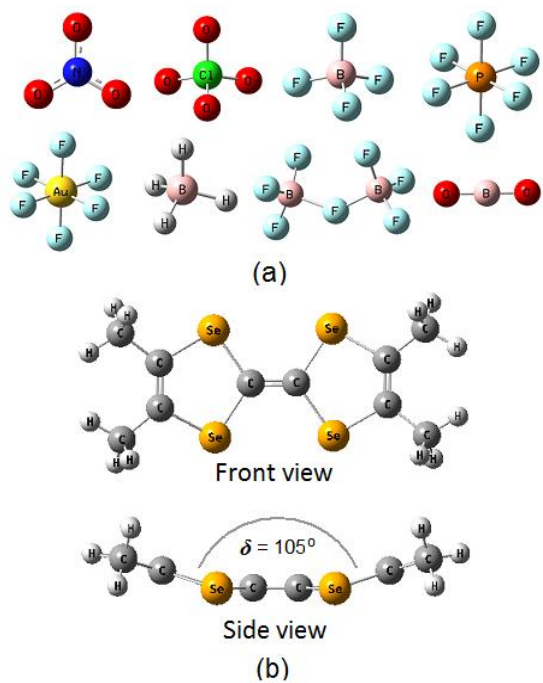

**Figure 1.** Equilibrium structures of superhalogen anions considered in this study (a) and TMTSF molecule (b).

The VDE of anions is calculated as the difference of total electronic energy of neutral system and corresponding anion both at anionic equilibrium structure.

[a] Dr. A. K. Srivastava, Prof. S. N. Tiwari
Department of Physics
DDU Gorakhpur University
Civil Lines, Gorakhpur, 273009, Uttar Pradesh, India
E-mail: ambrishphysics@gmail.com
[b] Mr. A. Kumar, Prof. N. Misra
Department of Physics
University of Lucknow
University Road, Lucknow, 226007, Uttar Pradesh, India

# RESEARCH PAPER

**Table 1.** Vertical detachment energy (VDE) of X⁻, interaction bond-length ($d_{int}$) and energy ($\Delta E_{int}$), net charge on X ($Q_X$), frontier orbital energy-gap ($E_{gap}$), dihedral angle between rings of TMTSF ($\delta$) and deformation energy ($\Delta E_{def}$) for various (TMTSF)$_2$X complexes.

| X | VDE (eV) | $d_{int}$ (Å) | $\Delta E_{int}$ (eV) | $Q_X$ (e) | $E_{gap}$ (eV) | $\delta$ | $\Delta E_{def}$ (eV) |
|---|---|---|---|---|---|---|---|
| NO$_3$ | 4.10 | 2.21-2.40 | 1.34 | -0.64 | 0.61 | 173° | 0.18 |
| BO$_2$ | 4.35 | 2.05 | 1.79 | -0.79 | 0.18 | 180° | 0.10 |
| BH$_4$ | 4.42 | 2.26-2.56 | 0.52 | -0.72 | 0.71 | 160° | 0.17 |
| ClO$_4$ | 5.65 | 2.34-2.44 | 2.57 | -0.86 | 0.68 | 158° | 0.15 |
| BF$_4$ | 7.39 | 2.31-2.36 | 4.62 | -0.71 | 0.71 | 159° | 0.17 |
| PF$_6$ | 8.22 | 2.33-2.74 | 5.11 | -0.87 | 0.23 | 173° | 0.12 |
| AuF$_6$ | 8.51 | 2.05-2.82 | 6.31 | -1.59 | 1.35 | 176° | 0.31 |
| B$_2$F$_7$ | 8.53 | 2.40-2.65 | 4.94 | -0.85 | 0.33 | 161° | 0.08 |

Table 1 lists the VDEs of these anions, which range from 4.10 eV for NO$_3^-$ to 8.22 eV for PF$_6^-$, and are large enough to suggest their superhalogen properties. Thus, the complex anions (X) in (TMTSF)$_2$X complexes belong to the class of superhalogen. Now, we consider some other superhalogen anions such as BO$_2^-$, BH$_4^-$, B$_2$F$_7^-$ and AuF$_6^-$ also displayed in Fig. 1. The VDEs of these superhalogen anions lie in the range 4.35 eV–8.53 eV as listed in Table 1. BO$_2^-$ is linear with bond lengths of 1.26 Å, same as those of NO$_3^-$. BH$_4^-$ is tetrahedral like BF$_4^-$ in which H atoms are employed instead of F as ligands. In B$_2$F$_7^-$, one of F ligands in BF$_4^-$ is substituted by BF$_4$ moiety itself. Like PF$_6^-$, AuF$_6^-$ is octahedral with the bond lengths of 1.96 Å. Therefore, these anions also possess some structural similarity with those of complex anions used in OSCs. We have also optimized the structure of TMTSF molecule as shown in Fig. 1. In TMTSF, there exist two pentagon rings containing two Se atoms each, which are connected head to head via C=C bond and substituted with two –CH$_3$ groups at other C positions. The ring systems are not coplanar, but possessing boat conformation having dihedral angle between rings ($\delta$) of 105° as marked in Fig. 1.

The equilibrium structures of (TMTSF)$_2$X complexes for aforementioned superhalogens (X) are displayed in Fig. 2. One can note that the geometrical structures of (TMTSF)$_2$X are significantly affected by the geometry of X. For trigonal planar NO$_3$, both TMTSF molecules possess stacked configuration. For tetrahedral BF$_4$ and ClO$_4$, both TMTSF units are almost parallel to each other. In case of octahedral PF$_6$, two TMTSF units become perpendicular to each other. Furthermore, TMTSF molecules arrange themselves parallel for tetrahedral BH$_4$ (like BF$_4$ and ClO$_4$) and perpendicular for octahedral AuF$_6$ (like PF$_6$). Moreover, these two units become collinear for linear BO$_2$ and stacked for B$_2$F$_7$ as clearly seen in Fig. 2.

Table 1 lists the interaction lengths ($d_{int}$) between TMTSF units and X along with their interaction energy ($\Delta E_{int}$), calculated as below:

$$\Delta E_{int} = E[X] + E[(TMTSF)_2] – E[(TMTSF)_2X],$$

where $E[..]$ represents total electronic energy of respective species. The interaction energy is an important parameter for measuring the strength of charge-transfer interaction in (TMTSF)$_2$X complexes. For typical OSCs, our calculated $\Delta E_{int}$ ranges from 1.34 eV for X = NO$_3$ to 5.11 eV for X = PF$_6$. In case of proposed (TMTSF)$_2$X complexes, this $\Delta E_{int}$ value varies between 0.52 eV for X = BH$_4$ and 6.31 eV for X = AuF$_6$. This may suggest that the stability of proposed complexes is comparable to or even greater than traditional OSCs, excluding the case of BH$_4$. Note that all these complexes are stabilized by hydrogen-bonding interactions expect (TMTSF)$_2$BH$_4$, which is stabilized by dihydrogen bonding.

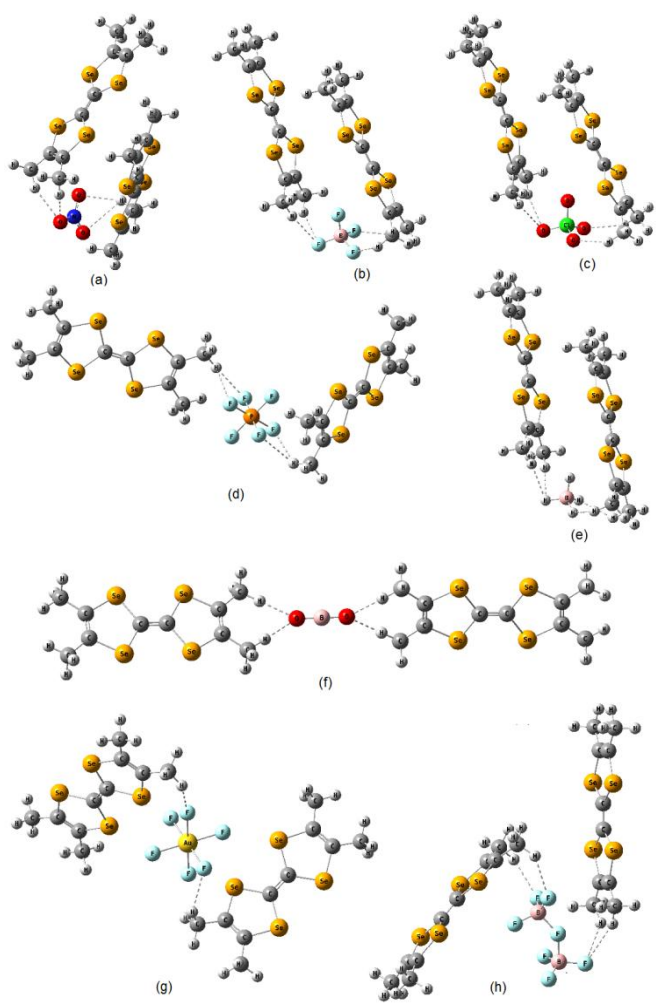

**Figure 2.** Equilibrium structures of (TMTSF)2X complexes at B3LYP/6-311++G(d,p) level of theory. The intermolecular interactions are shown by dashed lines.

As already known [13], the OSCs tend to have about half the donor (D) oxidized to D$^+$, with an average charge of +1/2 per D. In (TMTSF)$_2$X complexes, the electron transfer takes place from both TMTSF molecules to X moiety such that X becomes negatively charged leaving positive charge on TMTSF molecules. We have calculated atomic charge on X ($Q_X$) in these complexes and listed in Table 1. For traditional OSCs, the $Q_X$ value lies between -0.64e for X = $NO_3$ and -0.87e for X = $PF_6$. For other $(TMTSF)_2X$ complexes studied, the $Q_X$ takes values between -0.72e for X = $BH_4$ and -0.85e for X = $B_2F_7$, excluding $AuF_6$. Therefore, in the light of electron transfer and interaction energy, $BO_2$, $BH_4$ and $B_2F_7$ superhalogens are analogous to those of $NO_3$, $BF_4$, $ClO_4$, $PF_6$. For $AuF_6$, however, the electron transfer increases to -1.59e leading to significant increase in the interaction energy of $(TMTSF)_2AuF_6$ complex (see Table 1). In order to compare the reactivity (conductivity) of $(TMTSF)_2X$ complexes, we refer to their frontier orbital energy gap ($E_{gap}$) listed in Table 1. This energy gap corresponds to the band gap in solids. The $E_{gap}$ value of $(TMTSF)_2X$ complexes varies between 0.23 eV for X = $PF_6$ and 0.71 for X = $BF_4$. For X = $BO_2$, $BH_4$ and $B_2F_7$, this $E_{gap}$ value lies between 0.18 eV–0.71 eV. Evidently, $E_{gap}$ values of all $(TMTSF)_2X$ complexes studied are less than 1 eV, except that of $(TMTSF)_2AuF_6$ (1.35 eV). This may further suggest the applicability of $BO_2$, $BH_4$ and $B_2F_7$ superhalogens in the design of new OSCs.

As mentioned earlier, neutral TMTSF molecules possess boat like configurations (see Fig. 1). Due to electron-transfer from TMTSF molecules to X, their structures tend to be distorted and become planar. It has been suggested [14] that the conduction in OSCs leads to a coupling between electron-transfer and the boat deformation phonon modes and this electron-phonon coupling is responsible for the superconductivity. Therefore, we have analyzed the deformation of TMTSF boat structure towards planarity. In Table 1, we have also listed the dihedral angle between rings ($\delta$) of TMTSF molecules in $(TMTSF)_2X$ complexes and deformation energy ($\Delta E_{def}$) calculated as below:

$\Delta E_{def} = E_{deformed} − E_{boat}$

where $E_{boat}$ is total electronic energy of $(TMTSF)_2$ in their equilibrium boat configuration and $E_{deformed}$ is that of deformed $(TMTSF)_2$ configuration in $(TMTSF)_2X$ complex. This deformation creates destabilization in the TMTSF molecules, which can be quantified by $\Delta E_{def}$. Lower the $\Delta E_{def}$ value, higher the stabilization of neutral TMTSF in boat configuration. It has been established that the better OSCs possess larger stabilization of the boat deformation for neutral donor [13], hence smaller $\Delta E_{def}$ value. For OSCs considered here, the $\Delta E_{def}$ ranges 0.12–0.18 eV (2.8–4.2 kcal/mol). Note that $(TMTSF)_2PF_6$ is better OSC than $(TMTSF)_2NO_3$ in terms of critical temperature [3], which is in accordance with their smaller $\Delta E_{def}$ values. Therefore, considering $\Delta E_{def}$ as the most relevant parameter for OSCs, $(TMTSF)_2BO_2$ and $(TMTSF)_2B_2F_7$ may perform better than those of $(TMTSF)_2PF_6$ and $(TMTSF)_2ClO_4$. Although $(TMTSF)_2AuF_6$ possesses significantly larger $\Delta E_{def}$ value (0.31 eV) like $E_{gap}$ and $Q_X$ values, the $\Delta E_{def}$ of $(TMTSF)_2BH_4$ is also equal to that of $(TMTSF)_2BF_4$, consequently it should also possess desired properties for superconductivity.

In summary, the concept of superhalogen is indeed useful in designing potential candidates for OSCs. Having established that all acceptors of super-electrons in OSCs belong to the class of superhalogen, a new series of Bechgaard salts, $(TMTSF)_2X$ can be realized where X is a superhalogen. Similar conclusion also applies to Fabre salts as well as two-dimensional salts such as $(BETS)_2GaCl_4$ (BETS = bisethylenedithio tetraselenafulvalene) [15]. Our computed VDE of $GaCl_4^-$ is 6.25 eV, which clearly suggest its superhalogen nature. Our proposed $(TMTSF)_2BO_2$ and $(TMTSF)_2B_2F_7$ complexes are found to be suitable candidates for OSCs. These findings should motivate experimentalists for further exploration of their properties.

## Methods

We have used density functional theory (DFT) with B3LYP hybrid exchange-correlation functional [16] and triple-ζ basis set 6-311++G(d,p) including diffuse and polarization functions for all atoms (except Au, for which SDD pseudopotential has been employed) as implemented in Gaussian 09 program [17]. Considering the size of systems as well as popularity of the functional, the present computational scheme seems appropriate. The geometry of the structures were fully optimized without any symmetry constraint and followed by frequency calculations in order to ensure that the optimized structures belong to true minima in the potential energy surface.

## Acknowledgements

A. K. Srivastava acknowledges Science and Engineering Research Board (SERB), India for national post doctoral fellowship (NPDF) [Grant No. PDF/2016/001784].

**Keywords:** Bechgaard salts • density functional calculations • electron transfer • organic superconductors • superhalogens

[1]   W. A. Little, Phys. Rev. A 1964, 134, 1416-1424.
[2]   a) D. Jérome, A. Mazaud, M. Ribault, K. Bechgaard, J. Phys. Lett. 1980, 41, 95-98; b) N. Thorup, G. Rindorf, H. Soling, K. Bechgaard, Acta Cryst. B 1981, 37, 1236–1240.
[3]   a) K. Bechgaard, C. S. Jacobsen, K. Mortensen, H. J. Pedersen, N. Thorup, Solid State Commun.1980, 33, 1119-1125; b) S. S. P. Parkin, M. Ribault, D. Jerome and K. Bechgaard, J. Phys. C: Solid State Phys. 1981, 14, 5305-5326.
[4]   a) K. Bechgaard, K. Carneiro, M. Olsen, F. Rasmussen and C.S. Jacobsen, Phys. Rev. Lett. 1981, 46, 852; b) K. Bechgaard, K. Carneiro, F. B. Rasmussen, M. Olsen, G. Rindorf, C. S. Jacobsen, H. J. Pedersen, and J. C. Scott, J. Am. Chem. Soc.1981, 103, 2440-2442.
[5]   a) S. S. P. Parkin, E. M. Engler, R. R. Schumaker, R. Lagier, V. Y. Lee, J. C. Scott and R. L. Greene, Phys. Rev. Lett. 1983, 50, 270; b) J. M. Williams, H. H. Wang, M. A. Beno, T. G. Emge, L. M. Sowa, P. T. Copps, F. Behroozi, L. N. Hall, K. D. Carlson, and G. W. Crabtree, Inorg. Chem. 1984, 23, 3839; c) H. Urayama, H. Yamochi, G. Saito, S. Sato, T. Sugano, M. Kinoshita, A. Kawamoto, J. Tanaka, T. Inabe, T. Mori, Y. Maruyama, and H. Inokuchi, Synthetic Metals 1988, 27, A393.
[6]   a) A. F. Hebard, M. J. Rosseinsky, R. C. Haddon, D. W. Murphy, S. H. Glarum, T. T. M. Palstra, A. P. Ramirez, and A. R. Kortan, Nature 1991, 350, 600; b) A. Szasz, Journal Of Superconductivity 1993, 6, 99-106.
[7]   K. Tanigaki, T. W. Ebbessen, S. Saito, J. Mizuki, J. S. Tsai, Y. Kubo, S. Kuroshima, Nature 1991, 352, 222.
[8]   a) P. Winget, E. J. Weber, C. J. Cramer, D. G. Truhlar, Phys. Chem. Chem. Phys. 2000, 2, 1231; b) E. V. Patterson, C. J. Cramer, D. G. Truhlar, J. Am. Chem. Soc. 2001, 123, 2025; c) C. K. Kelly, C. J. Cramer, D. G. Truhlar, J. Phys. Chem. B 2007, 111, 408; d) M. H. Baik, R. A. Friesner, J. Phys. Chem. A 2002, 106, 7407; e) Y. Fu, L. Liu, Y. M. Wang, J. N. Li, T. Q. Yu, Q. X. Guo, J. Phys. Chem. A 2006, 110, 5874.
[9]   G. L. Gutsev and A. I. Boldyrev, Chem. Phys. 1981, 56, 277–283.

[10] X.-B. Wang, C.-F. Ding, L.-S. Wang, A. I. Boldyrev and J. Simons, J. Chem. Phys. 1999, 110, 4763.

[11] a) G. L. Gutsev and A. I. Boldyrev, Mol. Phys. 1984, 53, 23–31; b) G. L. Gutsev, J. Chem. Phys. 1993, 98, 444–452; c) G. L. Gutsev, R. J. Bartlett, A. I. Boldyrev and J. Simons, J. Chem. Phys. 1997, 107, 3867; d) X.-B. Wang and L.-S. Wang, J. Chem. Phys. 2000, 113, 10928–10933; e) M. Willis, M. Gotz, A. K. Kandalam, G. Gantefor and P. Jena, Angew. Chem. Int. Ed. 2010, 49, 8966; f) C. Sikorska and P. Skurski, Chem. Phys. Lett. 2012, 536, 34–38; g) A. K. Srivastava, N. Misra, New J. Chem. 2015, 39, 9543-9549.

[12] a) S. Giri, S. Bahera and P. Jena, Angew. Chem. Int. Ed. 2014, 53, 13916–13919; b) A. K. Srivastava and N. Misra, Electrochem. Commun. 2016, 68, 99–103; c) M. Czapla, I. Anusiewicz and P. Skurski, Chem. Phys. 2016, 465–466, 46–51.

[13] a) E. Demiralp, W. A. Goddard III, Synthetic Metals 1995, 72, 297-299; b) E. Demiralp, W. A. Goddard III, J. Phys. Chem. A 1997, 101, 8128-8131.

[14] E. Demiralp, S. Dasgupta, W. A. Goddard III, J. Am. Chem. Soc. 1995, 117, 8154-8158.

[15] K. Clark, A. Hassanien, S. Khan, K.-F. Braun, H. Tanaka and S.-W. Hla, Nature Nanotechnology 2010, 5, 261-265.

[16] a) A. D. Becke, Phys. Rev. A 1988, 38, 3098–3100; b) C. Lee, W. Yang and R. G. Parr, Phys. Rev. B 1988, 37, 785–789.

[17] M. J. Frisch, G. W. Trucks and H. B. Schlegel, et al., Gaussian 09, Rev. B.01 Gaussian Inc., Wallingford CT, 2010.